\documentclass{book}

\usepackage{graphics}
   \usepackage{graphicx}       \usepackage{bm}
    \usepackage{epsf}           \usepackage{epsfig}
\usepackage{amsmath}

\def\be{\begin{equation}}

\def\ee{\end{equation}}

\def\be{\begin{equation}}
  \def\ee{\end{equation}}
\def\bea{\begin{eqnarray}}
  \def\eea{\end{eqnarray}}

\begin{document}

\author{Alex W. Chin, Susana F. Huelga and Martin B. Plenio\\\\
Institut f\"{u}r Theoretische Physik,\\ Albert-Einstein-Allee 11,\\ Universit\"{a}t Ulm,\\ 89069 Ulm,\\ Germany}

\title{Chain representations of open quantum systems and their numerical simulation with time-adapative density matrix renormalisation group methods }

\author{Alex W. Chin\\\\
Institut f\"{u}r Theoretische Physik,\\ Albert-Einstein-Allee 11,\\ Universit\"{a}t Ulm,\\ 89069 Ulm,\\ Germany}
\author{Susana F. Huelga\\\\
Institut f\"{u}r Theoretische Physik,\\ Albert-Einstein-Allee 11,\\ Universit\"{a}t Ulm,\\ 89069 Ulm,\\ Germany}
\author{Alex W. Chin$^{1}$\\Susana F. Huelga$^{1}$\\Martin B. Plenio$^{1}$\\\\
$^{1}$Institut f\"{u}r Theoretische Physik,\\ Albert-Einstein-Allee 11,\\ Universit\"{a}t Ulm,\\ 89069 Ulm,\\ Germany}

\maketitle

\chapter{}
\section{Introduction}\label{intro}
As a result of uncontrollable interactions between quantum systems and their local environments, complex correlations develop between them which lead to the phenomena of \emph{decoherence} and \emph{relaxation} when only the quantum system is observed \cite{weiss, breuer2002theory, leggett,joos2003decoherence}. As almost no quantum states can ever be completely isolated from their surroundings, the dynamics of so-called \emph{open quantum systems} appear in almost all experiments in quantum physics, chemistry and biology, and a detailed understanding of the role of uncontrollable, \emph{noisy} environmental interactions is required to extract genuine quantum effects from realistic data.

In many cases, such as quantum optics and atomic physics, the effects of these processess are weak and relatively benign; while environmental interactions do degrade quantum effects, they do so on much slower timescales than those on which the effects operate and can be probed \cite{breuer2002theory,barnett1997methods,walls2008quantum}.  Under these conditions, these quantum effects cannot just be unambiguously observed, they can even be controlled and potentially \emph{harnessed} in new breeds of quantum device which can greatly outperform their classical analogues\cite{nielsen2002quantum}. Yet in other physical settings, such as the solid state and biological systems, the often strong and complex environmental interactions rapidly degrade quantum effects. Indeed, in many biological systems is has often been thought that relatively strong environmental noise is \emph{essential} for directing an essentially classical and irreversible - c.f. reversible unitary dynamics- migration of energy through the complex energy landscape which connects energy producing and energy consuming parts of the system \cite{van2000photosynthetic,may2004charge,blankenship2002molecular}. 

A good example of this latter paradigm is provided by pigment-protein complexes (PPCs) in photosynhetic organisms \cite{van2000photosynthetic,blankenship2002molecular}. 
These structures are involved in the early stages of light-harvesting and excitation energy transport (EET) which initiate the carbon-fixing reactions of photosynhesis. The wide variety of PPC structures share the common motif that they contain optically active \emph{chromophore} molecules embedded in a protein matrix which co-ordinates their spatial distribution.  In typical photosynthetic organisms, the PPCs are arranged so that particular complexes (antennae) absorb photons via the creation of electronic excitations (excitons) on their chromophores, whilst other complexes transport these excitations to reaction centers where electrons are released for photosynthetic chemistry \cite{blankenship2002molecular}. The passage of excitations from generation to consumption is generally achieved through the existence of energy gradients in the potential landscape of the inter- and intra-complex chromophores \cite{van2000photosynthetic,may2004charge,blankenship2002molecular} 
, which allows funneling of energy through dissipative processes induced by fluctuations of solvents and surrounding proteins.  Remarkably, for many photosynthetic systems under low light conditions, the quantum efficiency of photon capture, transport and charge generation is close to 100$\%$\cite{van2000photosynthetic,may2004charge,blankenship2002molecular}.

Although the migrating excitations in PPCs may be of a quantum mechanical nature,  it was normally assumed that the complex, high temperature environments of functioning PPCs would rapidly destroy inter-exciton coherences. Consequently, the dissipative funneling of energy could be intuitively described and understood by effectively classical rate-equation dynamics such as those provided by the  F\"orster and Dexter theories \cite{van2000photosynthetic,may2004charge}. 
 However, a much more complex picture of EET has recently emerged with the discovery of evidence for \emph{long-lasting inter-exciton coherences} in the EET dynamics of the Fenna-Matthews-Olson (FMO) complex \cite{gregory2007evidence}. This complex is extracted from green sulphur bacteria, and functions like a biomolecular `wire' that transports excitons from the light-harvesting chlorosomes to the charge-separating complex known as the reaction center \cite{blankenship2002molecular,muh2007,schmidt2010eighth}. Since the discovery of this evidence, similar effects have also been observed in complexes from marine algae and green plants \cite{collini2010coherently, calhoun2009quantum}, and further FMO experiments have now suggested coherence lifetimes of around $1.5$ ps at $77$ K and a few hundred femtoseconds at $277$ K \cite{panitchayangkoon2010long,hayes2010dynamics,caram2011extracting,hayes2011robustness,hayes2011extracting}. 

These inter-exciton coherence times are striking, as they are almost an order of magnitude longer than the coherence times of single excitonic transitions ($\sim100-200$ fs) \cite{hayes2010dynamics},  and as a result they persist over a significant fraction of the total transport time in typical PPCs \footnote{ The transport time for a single excitation to pass throuh the FMO complex is estimated to be $\sim5$ ps \cite{adolphs2006proteins}.}.  It has therefore been suggested that coherences may play an important role in driving the directed, highly-efficient EET observed in these complexes, and understanding this may provide valuable insights into how similar efficiencies could be achieved in artifical light-harvesting systems. However the mechanisms which preserve these coherences are currently unknown, and this and the intrincate interplay of noise and coherence that generates efficient transport has become a very rich and active problem \cite{nat2,plenio08,olaya2008efficiency, ishizaki2009theoretical,caruso09,caruso2010entanglement,chin2010noise,rebentrost2009environment,thorwart2009enhanced,fassioli2010distribution,sarovar2010quantum,jang2008theory,ishizaki2010quantum}.

The dynamical behaviour of interacting open quantum systems is frequently investigated in terms of simple dynamical models in which environmental
dephasing and relaxation are treated with Lindblad or Bloch-Redfield
master equations. These methods are both based on the assumptions
of weak system-bath coupling and the Markov approximation. However,
these approximations are not valid for many realistic systems, and assuming that the correlation time of the environments in these systems is much faster
than the system dynamics is frequently not justified. For instance, in typical
PPCs the dynamical timescales of the bath can be comparable or even slower
than
the EET dynamics \cite{ishizaki2009unified,ishizaki2010quantum,thorwart2009enhanced}.  Moreover, in the limit
of slow bath dynamics, perturbative treatments of the system-environment
coupling cannot be used even if the system-bath coupling is intrinsically
weak. Recently, important steps have been taken towards the development of non-perturbative and non-Markovian approaches, including generalised approximate master equations \cite{jang2008theory,fassioli2010quantum}, formally exact master equations which are unravelled by numerical hierarchy techniques (NHT) \cite{ishizaki2009unified,zhu2011modified}, stochastic methods \cite{roden2009influence}, and numerical path integral (NPI) techniques such as quantum monte carlo \cite{muhlbacher2008real} and QUAPI\cite{thorwart2009enhanced,nalbach2011iterative}. There are, however, limitations concerning the quality of the uncontrolled approximations made in some approaches \cite{jang2008theory,fassioli2010quantum, roden2009influence}, the restricted environmetnal structures accessible to several of these techniques  \cite{ishizaki2009unified,zhu2011modified}, and many of the numericlly-exact methods are expected to become less efficient with decreasing temperatures \cite{muhlbacher2008real,thorwart2009enhanced,ishizaki2009unified}. Given the detailed information about the real protein spectral
densities in PPCs is only just beginning to emerge \cite{olbrich2010quest}, a technique is required that can simulate EET for
arbitrary spectral densities and coupling strengths, thus allowing experiments carried out under different conditions, including low temperatures,
to be analyzed within one framework. 

The advanced numerical techniques such as NPI or NHT are approaches which deal with the time-evolution of the \emph{reduced} density matrix of the quantum sub-system.  The origins of the computational effort required to evaluate these schemes stems from the fact that without a separation of scales in PPC problems, the system and environment participate in the dynamics on an essentially equal footing. At a global level the dynamics has the character of a strongly-correlated many-body problem, suggesting an alternative approach to the problem based on numerical condensed matter theory methods. Because of the large number of environmental degrees of freedom, a direct simulation of the system and the environment appears rather daunting, but a number of powerful methods such numerical renormalisation group and sparse ploynomials space approaches have recently been developed to do precisely this \cite{bullarev08,anders07,alvermann09}. The key to the success of these methods is that the dynamics of the system-environment space can be accurately reproduced in a truncated Hilbert space, which is intimately related to the fact that many standard open-system Hamiltonians have an effectively $1D$ structure  which only contains nearest-neighbour interactions \cite{eisert2010colloquium}.  

In this chapter we introduce another many-body appoach to open quantum systems simulation that combines an exact analytical mapping of the problem onto an effective $1D$ nearest-neighbour model and the \emph{ time-adaptive density matrix renormalisation group} (t-DMRG) technique~\cite{schollwock2005density}.  Since its introduction the t-DMRG technqiue has proven to be one of the most powerful, accurate and versatile methods for simulating many-body dynamics in $1D$ \cite{schollwock2005density}, and in many cases leads to numerically exact results.  The mapping we use to generate the $1D$ representation also uses a novel application of the theory of orthogonal polynomials and a considerable portion of this chapter deals with this formalism and the \emph{physical} interpretation that this alternative picture provides for open-system dynamics. 

 This chapter is organised as follows: Section \ref{ham} introduces the standard open-system Hamiltonian and discusses the assumptions of this model. Section \ref{mapping} sets out in detail the formal mapping technique that generates an equivalent $1D$ representation of the open-system Hamiltonian which can be efficiently simulated by t-DMRG methods.  Section \ref{universal} points out a number of fundamental results on open-system structures which are revealed by this formal transformation, and points out how these might be used to increase the efficiency of future simulations. Section \ref{results} presents numerical examples of our mapping and t-DMRG approach, one of which points towards a novel mechanism for long-lasting excitonic coherence in PPCs. Finally, a set of conclusions and future prospects for this approach are given in Section \ref{conclusions}.  

\section{Open-system Hamiltonians and chain mappings}
\subsection{Standard model of open quantum system}\label{ham}

In this section we shall consider the most common model of an open quantum system, in which a quantum sub-system interacts with a macroscopic number of environmental degrees of freedom and the total sub-system and environment state evolves under a purely unitary dynamics. Dissipation and decoherence appear when the sub-system is observed without any knowledge of the state of the environment, leading to a non-unitary \emph{effective} dynamics for the sub-system's reduced density matrix. The total Hamiltonian can be written as $H=H_{s}+H_{I}+H_{B}$, where $H_{s}$ is the Hamiltonian of the quantum sub-system's degrees of freedom, $H_{B}$ is the free Hamiltonian of the environment and $H_{I}$ describes the interaction of the system and bath variables. For the typical problems described in the introduction, the quantum subsystem consists in a finite number  of quantum states which we denote as $|i\rangle$, and the system Hamiltonian can then be written in the general form
$$H_{s}=\sum_{i=1}^{N}\sum_{j=1}^{N}H_{ij} |i\rangle\langle j|,$$
where $H$ is a Hermitian matrix and $N$ is the total number of states which describe the quantum sub-system. For the excitation transport problems mentioned in Section \ref{intro}, it is natural to associate the states $|i\rangle$ with the presence of an excitation on a physical, spatially localised \emph{site}, in which case the diagonal matrix elements $H_{ii}$ give the local energies of these states and the off-diagonal matrix elements $H_{ij}$ quantify the probability amplitudes for these excitations to tunnel between sites $i$ and $j$.   

For pigment-protein complexes and an extremely wide range of systems encountered in physics, chemistry and biology, it is common to model the environment as a continuum of \emph{harmonic} oscillators which interact linearly with the operators of the system \cite{weiss, leggett,bullarev08, ishizaki2010quantum}. We shall represent such an oscillator environment in an explicit continuum representation \cite{bulla05,bullarev08}, which allows us to write $H_{I}$ as

\begin{equation}
H_{I}=\sum_{i=1}^{N}V_{si}\int_{0}^{1}dk h_{i}(k)(a_{i}(k)+a_{i}^{\dagger}(k))\label{hi}
\end{equation}
where $V_{si}$ are operators which act locally at site $i$ and $h_{i}(k)$ describes the coupling to field modes labelled by a continuous quantum number $k$. The modes are described by creation and annihilation operators $a_{i}^{\dagger}(k)$ and $a_{i}(k)$ respectively, which obey the bosonic commutation relation,
$$[a_{i}(k),a_{j}^{\dagger}(k')]=\delta_{ij}\delta(k-k').$$ 
We shall assume that $k$ lies within the finite interval $[0,1]$, leading to an enviornment with a sharp, finite bandwidth. Equation (\ref{hi}) is not the most general form of linear system-bath interaction, for instance the bath(s) could couple to multiple operators at each site, couple to collective modes of the sub-system or have a non-linear interaction in the bath variables. The restricted form we use is motivated by the usual assumption in EET problems that the primary effect of the environment is to induce fluctuations of the local site energies $H_{ii}$ \cite{ishizaki2010quantum}.  We have also assumed in Eq. (\ref{hi}) that operators on each system site couple to \emph{independent} (commuting) environments, and we will therefore not deal with the issue of spatially-correlated fluctuations\cite{hayes2010dynamics,nalbach2010quantum,fassioli2010quantum,nazir2009correlation,olbrich2010quest}. The free Hamiltonian of the oscillators is 
\begin{equation}
H_{B}=\int_{0}^{1}dk\,g_{i}(k)a_{i}^{\dagger}(k)a_{i}(k),
\end{equation}
where $g_{i}(k)$ is the dispersion of the field modes. The maximum frequency of the environment $\omega_{c}$ is given by $\omega_{c}=g(1)$.  The model is completed by specifying the spectral function of the environment $J(\omega)$, which in terms of the microscopic parameters of the Hamiltonian is given by \cite{bullarev08},
\begin{equation}
J(\omega)=h^2[g^{-1}(\omega)]\frac{dg^{-1}(\omega)}{d\omega}.
\label{jhg}
\end{equation}
In Eq. (\ref{jhg}), $g^{-1}(x)$ is the inverse function of the dispersion i.e. $g^{-1}(g(x))=x$. For all open-system problems where the environment is initially in a gaussian state, it can be shown rigourously that the influence of the environment on the reduced system dynamics is completely determined by $J(\omega)$ \emph{only} \cite{weiss,leggett, ishizaki2009unified, breuer2002theory}.  Equation (\ref{jhg}) provides a relation for obtaining $J(\omega)$ from a specific miscroscopic interaction model, however it is often the case that the spectral function itself is given or assumed, in which case the functions $h(k)$ and $g(k)$ are not uniquely specified. In the following section we will work with a fixed $J(\omega)$ and use the indeterminacy of $h(k)$ and $g(k)$ to effect the mapping we shall now present.         

\subsection{Unitary transformation of the environment}\label{mapping}
\begin{figure}
\includegraphics[width=12cm]{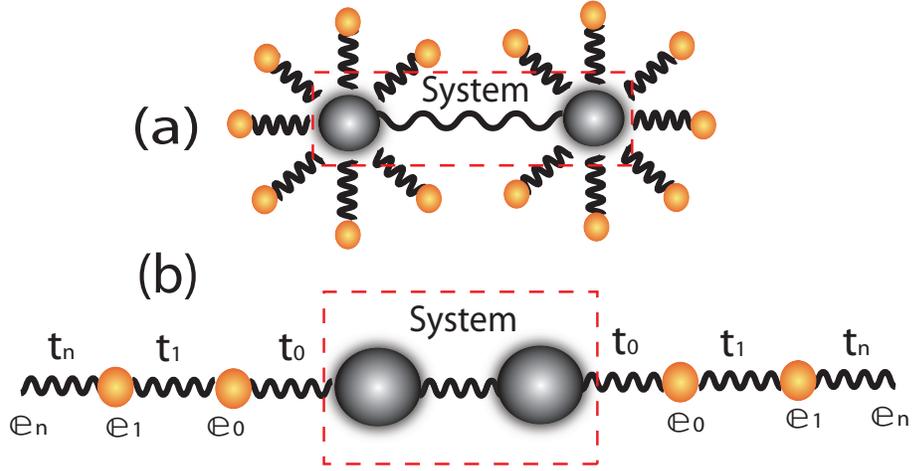}
\caption{ (a) Standard representation of a quantum dimer system in which each site is coupled to an independent continuum of harmonic oscillators. (b) After a unitary transformation of the oscillators, the entire system can be represented as a $1D$ chain with nearest neighbour interactions $t_{n}$ and local energies $\epsilon_{n}$. This equivalent many-body system can now be simulated efficiently using conventional t-DMRG techniques.  }
\label{chain}
\end{figure}
In this section we present the essential details of how we can convert the standard Hamiltonian structure of the  open quantum system shown in Fig. \ref{chain}.a into a 1-D form suitable for t-DMRG simulation.  The key insight is that the interaction of the quantum sub-system with all the modes of environmental oscillators is equivalent to the local interaction of the sub-system with one end of an infinite $1$D chain of coupled harmonic oscillators as shown in Fig.\ref{chain}.b. The existence of a chain representation of the environment has been  known for quite some time in a variety of quantum and classical contexts\cite{weiss,bullarev08,garg1985effect,hughes2009effectivei,hughes2009effectiveii,martinazzo2011communication}, and has been of particular use in the study of quantum impurity problems by numerical renormalisation group methods \cite{bullarev08,anders07}. In almost all previous approaches, the representation of the environment as a chain is used as an intermediate step that permits the application of a numerical technique. Consequently, the unitary transformation (see below) which maps the original open-system Hamiltonian onto a $1$D chain is often carried out numerically, following a discretisation of the continuous environmental spectrum to make the problem computationally tractable. However, these numerical mappings can often be numerical unstable, even for relatively unstructured environments.

In our approach we carry out the mapping \emph{formally}, using the theory of orthogonal polynomials to perform the mapping exactly and analytically. This formal approach allows us to make use of many of the rigourous results of orthogonal polynomial theory, and we shall show how their application reveals universal properties of open quantum systems which are independent of the specific forms of the environmental spectral function. Orthogonal polynomials also have rigourous connections to other important mathematical objects, such as continued fractions, Cauchy transforms and random matrices, and our theory provides a very general framework for investigating how these objects might also be applied to the problem at hand. 

A vast literature on orthogonal polynomials exists, and research into orthogonal systems is still extremely active, not least because of their important role in numerical quadrature, multi-dimensional interpolation, stochastic modelling, random matrices, approximation theory and analysis \cite{gautschi2004orthogonal,xiu2003wiener,deift2000orthogonal,mehta2004random,barthelmann2000high,baker1996pade}. A classic text is that of Szeg\"o \cite{szego1967orthogonal}, and many other fine books on the subject can be found in \cite{gautschi2004orthogonal,chihara1978introduction,ismail2005classical,nikiforov1991classical,askey1975orthogonal}.  For the most part, the material presented in this chapter only deals with the simplest types of orthogonal polynomial on the real line, and in what follows we shall use several standard results without proofs. The detailed proofs can be found in the any of the books above, but are also conveniently collected together in the context of the open quantum system problem in Chin, Rivas, Huelga and Plenio \cite{chinchain10}.

The starting point of the mapping is a unitary transformation which acts just on the environment oscillators. In order to prevent too many subscripts and summations from cluttering up our presentation, we shall only consider a single system site in what follows, dropping the site index $i$ throughout. As our open-system model consists of independent baths coupled to each site, the extension to multiple sites is trivial, and will be touched on again briefly in Section \ref{conclusions}. We implement the transformation by defining new bosonic modes according to 

\begin{equation}
b_{n}=\int_{0}^{1} dk h(k)\pi_{n}(k)\rho_{n}a(k),
\label{transa}
\end{equation}
where $h(k)$ is the coupling function in Eq. (\ref{hi}), $\pi_{n}(k)$ is a $n$th monic orthogonal polynomial (to be defined below), and $\rho_{n}$ is a normalisation constant. The corresponding transformation for $b_{n}^{\dagger}$ is obtained by taking the Hermitian conjugate of Eq. (\ref{transa}), and we note here that the parameters $h(k),\pi_{n}(k)$ and $\rho_{n}$ are all real-valued.  The function $\pi_{n}(k)$ is a monic $n$th degree polynomial $\pi_{n}(k)=\sum_{j=0}^{n}c_{nj}k^{j}$ where the monic condition means that $c_{nn}=1$. The coefficients of the polynomials $c_{j}$ are chosen so that they obey the following orthogonality condition, 

\begin{equation}
\int_{0}^{1}h^{2}(k)\pi_{n}(k)\pi_{m}(k)dk=\rho_{n}^{-2}\delta_{nm},
\label{ortho}
\end{equation}
which defines the normalization constant appearing in Eq. (\ref{transa}). The polynomials $\pi_{n}$ are known as monic orthogonal polynomials (MOPs) of the weight function $h^{2}(k)$. For a strictly positive weight function, as is manifestly the case for the weight function $h^{2}(k)$, a complete sequence of MOPs can always be found as a result of Favard's Theorem \cite{szego1967orthogonal, chinchain10}. The orthogonality condition immediately implies that,
\begin{eqnarray}
[b_{n},b_{m}^{\dagger}]&=&\rho_{n}\rho_{m}\int_{0}^{1}dk\int_{0}^{1}dk'h(k)h(k')\pi_{n}(k)\pi_{m}(k') [a(k),a^{\dagger}(k')]\nonumber\\
&=&\rho_{n}\rho_{m}\int_{0}^{1}dk h^{2}(k)\pi_{n}(k)\pi_{m}(k)\nonumber\\
&=&\delta_{nm}\label{unitary},
\end{eqnarray}
where we have used the commutation relation of the continuum field modes in the second line and the orthogoanlity relation of Eq. (\ref{ortho}) in the third. The transformation is real orthogonal and preserves the bosonic commuation relations of the new modes $b_{n}^{\dagger}$ \footnote{ We note that this transformation, and everything that follows in this section, would also hold true for an environment of \emph{fermionic} oscillators.}. The inverse transformation is given by, 
\begin{equation}
a(k)=\sum_{n=0}^{\infty}h(k)\rho_{n}\pi_{n}(k)b_{n},
\label{transb}
\end{equation}
which we now use to construct the chain Hamiltonian by substituting Eq. (\ref{transb}) into the original open-system Hamiltonian $H=H_{s}+H_{I}+H_{B}$. The transformation of the environment modes does not affect the system operators, and therefore $H_{s}$ and the system operator $V_{s}$ in the interaction term $H_{I}$  are unchanged by this operation. Let us now consider the effects of the transformation on the interaction Hamitonian $H_{I}$ and free bath Hamiltonian $H_{B}$ separately. The interaction term $H_{I}$ transforms in the following way, 
\begin{eqnarray}
H_{I}&=&V_{s}\int_{0}^{1}dkh(k)(a(k)+a^{\dagger}(k))\nonumber\\
&=&V_{s}\sum_{n=0}^{\infty}\rho_{n}(b_{n}+b^{\dagger}_{n})\int_{0}^{1}dkh^{2}(k)\pi_{n}(k)\nonumber\\
&=&V_{s}\sum_{n=0}^{\infty}\rho_{n}(b_{n}+b^{\dagger}_{n})\int_{0}^{1}dkh^{2}(k)\pi_{0}(k)\pi_{n}(k)\nonumber\\
&=&V_{s}\rho_{0}^{-1}(b_{0}+b_{0}^{\dagger}),
\end{eqnarray}
where we have used the fact that - by definition - $\pi_{0}(k)=1$, and then the orthogonality relation in the last line. The result of the transformation is that the system now couples to only a single mode $b_{0}$ of the new representation of the environment. We now turn to the bath Hamiltonian $H_{B}$. This transforms into, 
\begin{eqnarray}
H_{B}&=&\int_{0}^{1}dkg(k)a^{\dagger}(k)a(k)\nonumber\\
&=&\sum_{n=0}^{\infty}\sum_{m=0}^{\infty}b^{\dagger}_{n}b_{m}\int_{0}^{1}dk\,g(k)\pi_{n}(k)\pi_{m}(k)\label{hbg}.
\end{eqnarray}

At this point we cannot proceed further until the dispersion function $g(k)$ is specified. As discussed in Section \ref{ham}, the open-system dynamics of the sub-system is completely determined by the spectral function. Therefore, for a given $J(\omega)$ we have the freedom to \emph{choose} the form of either $h(k)$ or $g(k)$ as long as Eq. (\ref{jhg}) is respected. For reasons that will soon become apparent, we choose to take the dispersion to be $g(k)=\omega_{c}k$ \footnote{One can make other choices for the dispersion and this actually allows a number of different types of chain structures to be generated. The potential uses of these generalised structures in numerical applications is an interesting and open topic.}. This fixes $h^{2}(k)=\omega_{c}J(\omega_{c}k)$, and thus the MOPs of our transformation are orthogonal w.r.t. a weight function which is just proportional to the spectral function. The freedom to partition the spectral function between $h(k)$ and $g(k)$ is also used in the NRG approach\cite{bullarev08}, where it is used to logarithmically discretize the spectral function. The subsequent mapping onto a chain can also be solved analytically with generalised MOPs, and an example of such a solution is given in section \ref{log}.  The linear form of $g(x)$ now allows us to use another general property of MOPs, which is that they all obey the following three-term recurrence relation \cite{szego1967orthogonal,chinchain10},

\begin{equation} 
k\pi_{n}(k) =\alpha_{n}\pi_{n}(k)+\beta_{n}\pi_{n-1}(k)+\pi_{n+1}(k), \hspace{1cm} \pi_{-1}(k)=0,
\end{equation}
where the sequence of numbers $\alpha_{n},\beta_{n}$ are unique for a given weight function and are given by $\alpha_{n}=\rho_{n}^{2}\int_{0}^{1}dk h^{2}(k)\,k\pi_{n}(k)\pi_{n}(k)$ and $\beta_{n}=\rho_{n}\rho_{n+1}\int_{0}^{1}dk\,h^{2}(k)k\pi_{n}(k)\pi_{n-1}(k) $. If we now substitute the linear form of $g(k)$ into Eq. (\ref{hbg}) and use the recurrence and orthogonality relations, we obtain

\begin{eqnarray}
H_{B}&=&\sum_{n=0}^{\infty}\sum_{m=0}^{\infty}b^{\dagger}_{n}b_{m}\int_{0}^{1}dk\,\omega_{c}k\pi_{n}(k)\pi_{m}(k),\nonumber\\
&=&\sum_{n=0}^{\infty}\sum_{m=0}^{\infty}b^{\dagger}_{n}b_{m}\int_{0}^{1}dk\,\pi_{n}(k)(\alpha_{m}\pi_{m}+\beta_{m}\pi_{m-1}(k)+\pi_{m+1}(k)),\nonumber\\
&=&\omega_{c}\sum_{n=0}^{\infty}\left(\alpha_{n}b_{n}^{\dagger}b_{n}+\frac{\rho_{n+1}\beta_{n+1}}{\rho_{n}}b_{n}^{\dagger}b_{n+1}+\frac{\rho_{n}}{\rho_{n+1}}b_{n+1}^{\dagger}b_{n}\right).
\end{eqnarray}

Due to the choice of the linear dispersion, the transformed bath Hamiltonian takes the form of a one dimensional harmonic chain with only nearest neighbour coupling. From the definitions of $\beta_{n}$ and $\rho_n$ one can easily show that $\rho_{n}/\rho_{n+1}=\sqrt{\beta_{n+1}}$, allowing us to rewrite $H_{B}$ in the final, symmetrised form,

\begin{equation}
H_{B}=\sum_{n=0}^{\infty}\epsilon_{n}b_{n}^{\dagger}b_{n}+t_{n}b_{n}^{\dagger}b_{n+1}+t_{n}b_{n+1}^{\dagger}b_{n},
\end{equation}
where $\epsilon_{n}=\omega_{c}\alpha_{n}$ and $t_{n}=\omega_{c}\sqrt{\beta_{n+1}}$. We have now completed the formally exact transformation from the original Hamiltonian to a $1-D$ nearest-neighbour Hamiltonian. Collecting together all the transformed terms, the total Hamiltonian in the chain representation is given by,

\begin{equation}
H_{total}=H_{s}+\eta V_{s}(b_{0}+b_{0}^{\dagger})+\sum_{n=0}^{\infty}\epsilon_{n}b_{n}^{\dagger}b_{n}+t_{n}b_{n}^{\dagger}b_{n+1}+t_{n}b_{n+1}^{\dagger}b_{n},
\end{equation}
where using $h^{2}(k)=\omega_{c}J(\omega_{c} k)$ we have defined the coupling constant $\eta$,

\begin{equation}
\eta^2=\rho_{0}^{-2}=\int_{0}^{1}dk\,h^{2}(k)=\int_{0}^{\omega_{c}}d\omega J(\omega).
\end{equation}

The dynamics of the many-body system and bath state under this  Hamiltonian structure can now simulated using t-DMRG, as will be described in Section \ref{results}. However before presenting the simulation technique we shall briefly describe some physical implications of the exact mapping.

\subsection{Universal properties of continuous environments and the determination of the chain frequencies and couplings}\label{universal}

\begin{figure}
\includegraphics[width=12cm]{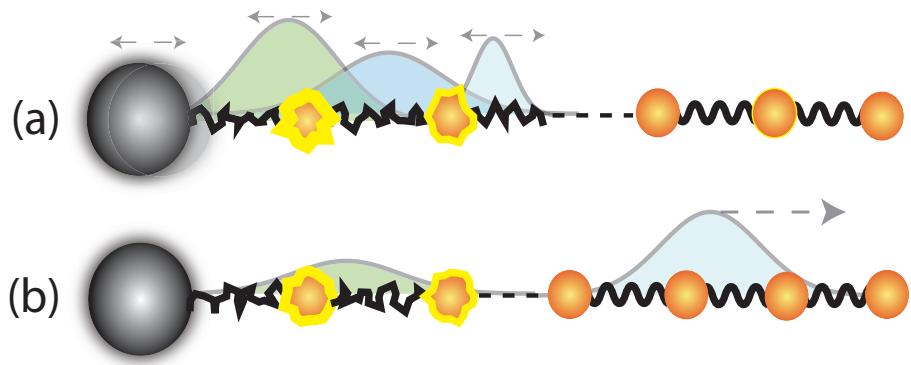}
\caption{Illustrative sketch of open-system dynamics in the chain representation. (a) Sub-system initially injects excitations (shown as wavepackets) into inhomogeneous region of the chain. Scattering from inhomogeneity causes back action of excitations on the system at later times and leads to memory effects and non-Markovian sub-system dynamics. (b) At long times, after multiple scattering, excitations penetrate into the homogeneous region and propagate away from the system without backscattering. This leads to irreversible and Markovian excitation absorption by the environment. }
\label{memory}
\end{figure}

In section \ref{mapping} we derived the relation between the chain oscillator frequencies $\epsilon_{n}$ and the couplings $t_{n}$ to the MOPs recurrence coefficients $\alpha_{n}$ and $\beta_{n}$. The determination of the chain which corresponds to an environment characterised by a given $J(\omega)$ therefore reduces to the problem of determining the recurrence coefficients of the MOPs w.r.t. the weight function $J(\omega)$. For several important weight functions, these recurrence coefficients can be given by simple analytical formula.  A comprehensive list and analysis of these classical MOPs can be found in \cite{gautschi2004orthogonal,szego1967orthogonal,chihara1978introduction,ismail2005classical,nikiforov1991classical,askey1975orthogonal}.  A very useful example is provided by the shifted-Jacobi polynomials $P_{n}^{0,s}(k)$, which are defined on the interval $k\in [0,1]$. These polynomials are orthogonal w.r.t. the Caldeira-Leggett spectral density/weight function $J(\omega_{c} k)=\alpha \omega_{c}^{1-s}\omega^{s}\theta(\omega_{c}-\omega)$, which is often used in discussions of the spin-boson model, the archetypical model of an open quantum system \cite{weiss, leggett, chin06,bullarev08,anders07,alvermann09,nalbach2010ultraslow}. The corresponding chain frequencies, inter-chain couplings and coupling to the quantum system are given by,

\begin{eqnarray}
\epsilon_{n}&=&\frac{\omega_{c}}{2}\left(1 + \frac{s^{2}}{(s + 2n) (2 + s + 2 n)}\right)\label{exenergies},\\
\nonumber\\
t_{n}&=&\frac{\omega_{c}(1 + n) (1 + s + n) }{(s + 2 +2n) (3 + s + 2n)}\sqrt{\frac{3 + s + 2n }{ 1 + s + 2 n}},\label{extn}\\
\eta^2&=&\frac{\alpha\omega_{c}}{1+s}.\label{etas}
\end{eqnarray}
As one can see from Eqs. (\ref{exenergies}), (\ref{extn}) and (\ref{etas}), the energy scale of the bath Hamiltonian and interaction terms are set by $\omega_{c}$ (as one would expect) and the total dynamics of the open quantum system are determined by the dimensional-less parameters $\alpha$ and the eigenvalues of $H_{s}/\omega_{c}$. In addition to this, we can also immediately infer the asymptotic ($n\rightarrow \infty$) parameters of the chain, finding that $\epsilon_{n}\rightarrow \omega_{c}/2$ and $t_{n}\rightarrow \omega_{c}/4$ as $n\rightarrow \infty$. These asymptotic values do not depend on the values of $s$ which characterise the shape of $J(\omega)$ and are thus \emph{universal} for spectral densities of the Jacobi power-law form.  At large distances from the sub-system the harmonic chain becomes homogenous and excitations in this part of the chain become simple harmonic waves. Using the asymptotic values of $\epsilon_{n}$ and $t_{n}$ one can simply diagonalise the homogenous part of the chain, yielding the dispersion $\Omega(q)=\frac{1}{2}\omega_{c}(1-\cos(\pi q)) $ of excitations with wavevector $q$. As shown in Fig. \ref{memory}, the asymptotic region of the chain can be loosely thought of as a type of `transmission' line, whose homogeneity ensures no backscattering of excitations towards the system. As sketched in Fig. (\ref{memory}), this enables this region to carry away excitations from the sub-system irreversibily at long times, as one would expect for a dissipative environment.  On physical grounds we would also require that this region of the chain should be able to support excitations at all frequencies covered by the original spectral function, and indeed it can be seen that the asymptotic values $\epsilon_{n}$ and $t_{n}$ are the only values which give the correct bandwidth for the asymptotic region. 

The emergence of a universal asymptotic chain appears directly from the analytical formula for the Jacobi recurrence coefficients, and can be physically motived by the arguments given above. Indeed, on the basis of the physical arguments one might expect this asymptotic  homogeneity to appear for \emph{any} finite bandwidth environment, and this indeed turns out to be the case. The proof is due to Szeg\"o\footnote{We have rephrased the theorem in terms of the chain parameters. The theorem presented in Refs. \cite{szego1967orthogonal, chinchain10} is actually a statment about the asymptotic values of the recurrence coefficients of a sequence of orthogonal polynomials defined over a finite interval. }, who was able to show that the asymptotic values for $\epsilon_{n}$ and $t_{n}$ are, respectively, $\omega_{c}/2$ and $\omega_{c}/4$ for \emph{any} weight function $h^{2}(k)$ which obeys the inequality \cite{szego1967orthogonal,chinchain10}

\begin{equation}\label{SzegoCondition}
\int_{0}^1\frac{\ln h^{2}(k)}{\sqrt{1-(2k-1)^2}}dk>-\infty.
\end{equation}

Weight functions which obey Eq. (\ref{SzegoCondition}) are said to belong to the Szeg\"o class \cite{chihara1978introduction,gautschi2004orthogonal}. In the context of the open-system problems we are considering, a huge range of spectral functions over a finite interval falls within the  Szeg\"o class and thus the homogenous asymptotic chain appears in almost every chain representation of a physical, finite bandwidth environment. Notable example of non-Szeg\"o spectral function corresponds to spectral functions containing band gaps or spectral functions defined over semi-infinite domains. We shall not consider these cases in this chapter, but they are dealt with in Chin, Rivas, Huelga and Plenio \cite{chinchain10}.

The existence of a universal asymptotic form of the chain region leads to a very appealing and simple picture of memory effects and non-Markovianity in open-system dynamics. The chain structure itself implies a natural causality, or set of timescales, over which different regions of the chain contribute to the dynamics as shown in Fig. (\ref{memory}). At early times the system interacts with the modes on the left of the chain, injecting excitations into this region which then begin to propagate to the right. Because of the inhomogeniety of this region, which is dependent on the specific form of the spectral function, these excitations will undergo scattering and some of them will return and act on the system at a later time. These backscattering proceses represent memory effects in the system-bath interaction and depend sensitively on the form of the spectral function. At later times excitations propagate into the homogeneous asymptotic region of the chain and are effectively absorbed irreversibly by the environment. The dynamics of this process is independent of the shape of the spectral function and describe a dissipative, long-time Markovian dynamics of the sub-system. Therefore in the chain representation the bath corrrelation time and related memory effects are associated with the typical time it takes an excitation to exit the inhomogenous region close to the system. This time depends on the form of the bath  which determines the size and spatial extent of the backscattering potential seen by these excitations. The strengh of non-Markovian effects on the sub-system dynamics depends on how excited the inhomogenous region is during the time evolution and will therefore depend on the rate at which excitations are injected into this region i.e. it will be dependent on the coupling strength.  Non-trivial, initially non-Markovian dynamics is therefore expected when the excitation injection rate is much larger than the rate of escape from the inhomogenous region of the chain at early times. 

\begin{figure}
\includegraphics[width=12cm]{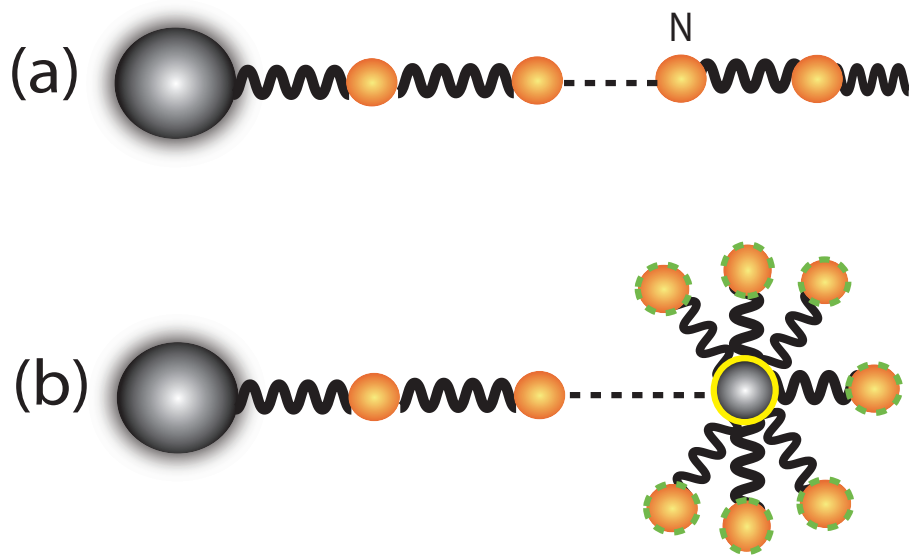}
\caption{(a) Diagonalising the homogeneous part of the chain after site $N$ leads to an effective environment acting on this terminal oscillator, as shown in (b). This terminal spectral density is universal for any spectral density in the Szeg\"o class, suggesting that complex environments may be efficiently handled by only treating the initial oscillators of the chain which encode the specific characteristics of a given environment.  }
\label{terminal}
\end{figure}

Another representation of this idea of a non-trivial, non-universal early time dynamics which evolves into a universal dissipative dynamics is shown in Fig. (\ref{terminal}). After the chain parameters have, to within some arbitrary tolerance, converged to the asymptotic values at site $N$, the remainder of the chain to the right is diagonalised. This provides an effective environment acting on the $N$th member of the chain which in the limit $N\rightarrow\infty$  possesses a universal spectral function proportional to the Wigner semi-circle distribution $J_{T}(\omega)\propto\sqrt{\omega(\omega_{c}-\omega)}$, which is an important equilibrium distribution in random matrix theory and which is also imitately related to the properties of Chebyshev polynomials \cite{chinchain10,mehta2004random,gautschi2004orthogonal}.  This representation suggests a possible reduction in the complexity of simulating the dynamics of a complex environment, as in many cases the convergence of the chain parameters is rather rapid\footnote{For the Jacobi spectral functions the parameters converge to their asymptotic values as $\frac{s^2}{n^{2}}$ as $n\rightarrow\infty$.}. It may therefore be possible to simulate the system by treating only the first few non-trivial sites of the chain explicitly, and then using numerically cheaper semi-classical, or even classical,  methods to model the damped mode at site $N$. 

As t-DMRG simulates the entire wavefunction of the system and environment, we will be able to explore the correlations and entanglement between the system and bath, allowing us to accurately assess the quality of such an approximation and how to improve upon it systematically. Investigating system-bath correlations may also be of some relevance for understanding how entanglement is generated between different components of open systems \cite{caruso09,caruso2010entanglement, sarovar2010quantum}, and is of direct relevance for the recently-developed theory of measures of non-markovanity \cite{rivas2010entanglement,breuer2009measure}. An important practical application of having access to bath information is that we can also explore at the microscopic level how preparation and propagation of \emph{wave packet dynamics} in complex environments can influence EET networks.  

This idea of the reduction of complex environmental spectra has also been addressed by Burghardt et al.\cite{hughes2009effectivei,hughes2009effectiveii,martinazzo2011communication}, who have derived an iterative formula for the effective spectral density acting on site $N$ as $N$ is increased.  Using a mass-weighted co-ordinate representation of the environment and chain,  Martinazzo et al.  also empirically found that the spectral density converges to a universal limit under certain conditions, and that this terminal spectral density has the Ohmic Rubin model form $J_{T}(\omega)\propto \omega\sqrt{1-\frac{\omega^2}{\omega_{c}^{2}}}$ \cite{martinazzo2011communication}.  Their method makes extensive use of continued fractions and Cauchy tranforms, which are intimately related, via the Jacobi matrix, to orthogonal polynomials \cite{gautschi2004orthogonal,  szego1967orthogonal,chihara1978introduction}.  As shown in Weiss \cite{weiss}, the Rubin spectral density can also be represented by a coupling to a \emph{uniform} chain of harmonic oscillators coupled by nearest neighbour interactions, and the formal links between these approaches is currently being investigated within the framework of orthogonal polynomial theory.

\subsection{ Continuous, discrete and mixed spectral densities} \label{dis}

In the previous section we dealt with chain representations related to continuous spectral functions over a finite interval. In many situations we also encounter spectral densities containing discrete contributions, either as a result of the physical presence of strong coupling to discrete modes of the environment or an artifical discretisation of the environment that has been performed to facilitate a numerical approach to the problem. As discussed in Refs. \cite{gautschi2004orthogonal, szego1967orthogonal,chihara1978introduction, nikiforov1991classical,ismail2005classical,chinchain10} it turns out that MOPs can also be found for such spectral functions, permitting the formal transformation of these problems into the $1D$ harmonic chain problem. We shall illustrate this with analytical results for the important case of a logarithmically-discretized power-law spectral density. This artifically discretised spectral density plays an important role in the powerful numerical renormalisation group approaches to quantum impurity problems \cite{bullarev08}.  Numerical results for a physical spectral density with a discrete component will be presented in Section \ref{ARspec}. 

To handle discrete components we consider spectral functions $J(k)$ of the form,
\begin{equation}
J(k)=h^{2}(k)+\sum_{j=1}^{N}w_{j}\delta(k-k_{j}) \hspace{1cm} k,k_{j}\in[0,1],\label{mixed}
\end{equation}
where $h^{2}(k)$ is a continuous, non-negative spectral density, $w_{j}$ are positive weights for discrete contributions to the sepctral density and $k_{j}$ the (scaled) frequencies at which these discrete features occur. Under these conditions it can be shown that a set of MOPs can always be found which obey \cite{gautschi2004orthogonal,gautschi2005orthogonal},
\begin{eqnarray}
\int_{0}^{1}dk J(k)\pi_{n}(k)\pi_{m}(k)&=&\int_{0}^{1}dkh^{2}(k)\pi_{n}(k)\pi_{m}(k)+\sum_{j=1}^{N}w_{j}\pi_{n}(k_{j})\pi_{m}(k_{j})\nonumber\\
&=&\rho_{n}^{-2}\delta_{nm},
\end{eqnarray}
and that these MOPs possess the key properties we need to implement the chain transformation, such as the three-term recurrence relation. In the extreme case where all $w_{j}$'s are zero, an infinite sequence of MOPs, like those we have already considered, arises. In the opposite extreme where $h^{2}(k)=0$, there is a finite number $N$  of discrete MOPs which obey the discrete orthogonality condition $\sum_{j=1}^{N}w_{j}\pi_{n}(k_{j})\pi_{m}(k_{j})=\delta_{nm}$. Just like in the continuous case, there exists a number of classical discrete MOPs whose properties can be expressed in analytical form, and a comprehensive list can be found in \cite{nikiforov1991classical}. In the mixed case the sequence of MOPs is also infinite, and while a few special cases can be solved analytically \cite{gautschi2004orthogonal,chihara1978introduction}, the MOPs for these cases normally have to be found numerically.  

For the general mixed spectral density of Eq. (\ref{mixed}) a number of very efficient algorithms have been developed for computing the values of the recurrence coefficients $\alpha_{n},\beta_{n}$ which enter the chain Hamiltonian. The most effective of these for mixed problems involve adaptable discretisation and quadrature schemes which are collected in W. Gautschi's software package ORTHOPOL \cite{gautschi2005orthogonal,gautschi2004orthogonal}. These algorithms were used to determine the chain parameters for the numerical t-DMRG results in Section \ref{results}.

Before presenting numerical simulations we shall quickly give a practically useful example of an analytical solution to an important and purely discrete MOP problem.    

\subsection{Logarithmically-discretised spectral density}\label{log}
In numerical renormalisation group (NRG) studies of quantum impurity problems of the spin-boson model-type \cite{bulla05,bullarev08,anders07}, a Hamiltonian of the form of Eq. (\ref{ham}) is first logarithmically discretized following the procedure set out in \cite{bullarev08}. The Hamiltonian $H$ after the logarithmic discretisation of the reservoir continuum takes the discrete form
\begin{equation*}
H=H_{s}+\frac{V_{s}}{2\sqrt{\pi}}\sum_{n=0}^{\infty}\gamma_{n}(a_{n}+a_{n}^{\dagger})+\sum_{n=0}^{\infty}\zeta_{n}a_{n}^{\dagger}a_{n}\label{hlsb},
\end{equation*}
where
\begin{eqnarray}
\gamma_{n}^{2}&=&\frac{2\pi\alpha}{1+s}\omega_{c}^{2}(1-\Delta^{-(1+s)})\Delta^{-n(1+s)},\\
\zeta_{n}&=&\frac{s+1}{s+2}\frac{1-\Delta^{-(s+2)}}{1-\Delta^{-(s+1)}}\omega_{c}\Delta^{-n}.
\end{eqnarray}
It has been shown by Bulla et al. that this Hamiltonian can then be mapped to a nearest-neighbour chain Hamiltonian of the form \cite{bulla05}, 

\begin{equation}
H_{c}=H_{s}+\frac{1}{2}\sqrt{\frac{\eta_{0}}{\pi}} V_{s} (b_{0}+b_{0}^{\dagger})+\sum_{n}\omega_{n}b_{n}^{\dagger}b_{n}+t_{n}b_{n+1}^{\dagger}b_{n}+t_{n}b_{n}^{\dagger}b_{n+1}\label{hcl},
\end{equation}
by a real orthogonal tranformation $b_{n}^{\dagger}=\sum_{m}U_{nm}a_{m}^{\dagger}$ provided that the matrix elements $U_{nm}$ obey the three-term recurrence relation,
\begin{equation}
\zeta_{n}U_{mn}=\omega_{m}U_{mn}+t_{m}U_{m+1n}+t_{m-1}U_{m-1n}.\label{rec}
\end{equation}
In the NRG approach, this recurrence relation is solved numerically by a simple iterative procedure that rapidly becomes unstable as the size of the chain increases. However, the appearance of a real symmetric three-term recurrence relation suggests that a closed form solution exists in terms of suitably chosen orthogonal polynomials. The resulting polynomials are in fact well characterised, allowing us to find the chain parameters of the logarithmically-discretised chain exactly.  These polynomials are the \emph{little-q Jacobi polynomials} $p_{n}(\Delta^{-m},\Delta^{-s},1|\Delta^{-1})$. These are normally not part of the classical scheme of discrete orthogonal polynomials and are in fact \emph{q-analogues} of the classical Jacobi polynomials. A detailed discussion and a list of the other known \emph{q-orthogonal polynomials} can be found in \cite{koekoek1996askey}. Their important properties for our purposes is that that they obey the orthogonality relation, 
\begin{eqnarray}
\delta_{nm}N^{2}_{n}&=&\sum_{k=0}^{\infty}\Delta^{-k(1+s)}p_{n}(\Delta^{-k},\Delta^{-s},1|\Delta^{-1})p_{m}(\Delta^{-k},\Delta^{-s},1|\Delta^{-1}),
\label{ortho1}
\end{eqnarray}
and the recurrence relation
\begin{eqnarray}
\Delta^{-n}p_{j}(\Delta^{-n},\Delta^{-s},1|\Delta^{-1})&=&(A_{j}+C_{j})p_{j}(\Delta^{-n},\Delta^{-s},1|\Delta^{-1})-A_{j}p_{j+1}(\Delta^{-n},\Delta^{-s},1|\Delta^{-1})\nonumber\\
&-&C_{j}p_{j-1}(\Delta^{-n},\Delta^{-s},1|\Delta^{-1}).
\end{eqnarray}
The normalisation constants $N_{n}$ in Eq. (\ref{ortho1}) and the recurrence constants $A_{n},B_{n}$ and $C_{n}$ can be expressed in closed form and can be evaluated easily without any need for potentially unstable iterative techniques. The various coefficients are listed in \cite{chinchain10,koekoek1996askey}. As shown in Ref. \cite{chinchain10}, with just these two properties one can prove that the unitary matrix,
\begin{eqnarray}
U_{nm}&=&\frac{\Delta^{-\frac{m(1+s)}{2}}p_{n}(\Delta^{-m},\Delta^{-s},1|\Delta^{-1})}{N_{n}},\label{jac}
\end{eqnarray}
solves the recurrence relation of Bulla, and thus carries out the mapping exactly. The resulting chain parameters of Eq. (\ref{hcl}) are then given by,
\begin{eqnarray}
\omega_{n}&=&\zeta_{s}(A_{n}+C_{n}),\label{e}\\
t_{n}&=&-\zeta_{s}\left(\frac{N_{n+1}}{N_{n}}.\right)A_{n}.\label{t}
\end{eqnarray}

\section{Numerical results and applications}\label{results}
We now demonstrate the implementation of our joint mapping and t-DMRG approach with some specific spectral densities of relevance for PPCs in
photosynthetic organisms. Although the richly structured environments used in the PPC literature are taken as challenging examples, it should be emphasized that this new simulation
tool is completely general, and can be applied to any system linearly coupled to bosonic or fermionic environments of arbitrary spectral density. The PPC results were first presented in Prior et al. \cite{prior10}. We shall consider a dimer system consisting of two sites $1$ and $2$ with local site energies $\epsilon_{1},\epsilon_{2}$, and which are connected by a tunneling amplitude $J$. Each site is coupled to its own independent environment as in Eq. (\ref{hi}), and each environment is described by identical spectral densities. After the chain transformation, the system structure is exactly as shown in Fig. (\ref{chain})b. The initial state of the system for all simulations is taken as the separable pure state $\rho=\rho_{s}\otimes\rho_{B}$, where $\rho_{s}$ describes an initial excitation on site $1$ and $\rho_{B}$ is the vacuum state for the chain. The pure state initial condition implies that we are considering the open-system dynamics at zero temperature.These conditions on the spectral densities and states were chosen for simplicity and for their correspondence to the physical conditions found in PPCs immediately after photoexcitation, but these conditions are not required for the successful implementation of our method. 

The pure state t-DMRG algorithm employed is the standard one presented in Refs. \cite{schollwock2005density,vidal04,white2004real,daley2004time}, which is used to evolve the total wavefunction $|\Psi(t)\rangle$ of the dimer and environments in real time. Observables of the sub-system $O_{s}$ at time $t$ were obtained from the expectation values $\langle O_{s}\rangle (t)=\langle \Psi(t)|O_{s}|\Psi(t)\rangle$. In all t-DMRG simulations, we found that the results converged to less than $0.1\%$ with just $11$ bosonic levels per site, $30$ Schmidt coefficients and $100$ chain sites over the whole dynamics \cite{prior10}. 

\subsection{ The overdamped Brownian oscillator spectral density}\label{odbo}

 To start with, we look at the overdamped Brownian
oscillator spectral density which has been extensively studied in the context of PPC dynamics, and which can be solved in an numerically exact way in the high-temperature limit \cite{ishizaki2009unified}.
In our notation,
the overdamped Brownian oscillator spectral density has a simple Ohmic form,
\begin{equation}
J(\omega)=\frac{8\lambda\gamma\omega}{\omega^{2}+\gamma^2},
\label{jobo}
\end{equation}
where $\lambda$ is the reorganisation energy of the bath, defined by $\lambda=\frac{1}{4\pi}\int_{0}^{\omega_{c}}J(\omega)\omega^{-1}d\omega$, and is taken as our measure of the site-environment coupling strength. The parameter $\gamma$ approximately sets the dynamical response time of the bath, and the following simulations use values of $\gamma$ which are smaller than the dimer energy scales in order to observe non-Markovian effects \cite{ishizaki2009theoretical,thorwart2009enhanced, roden2009influence}. 

\begin{figure}
\includegraphics[width=12cm]{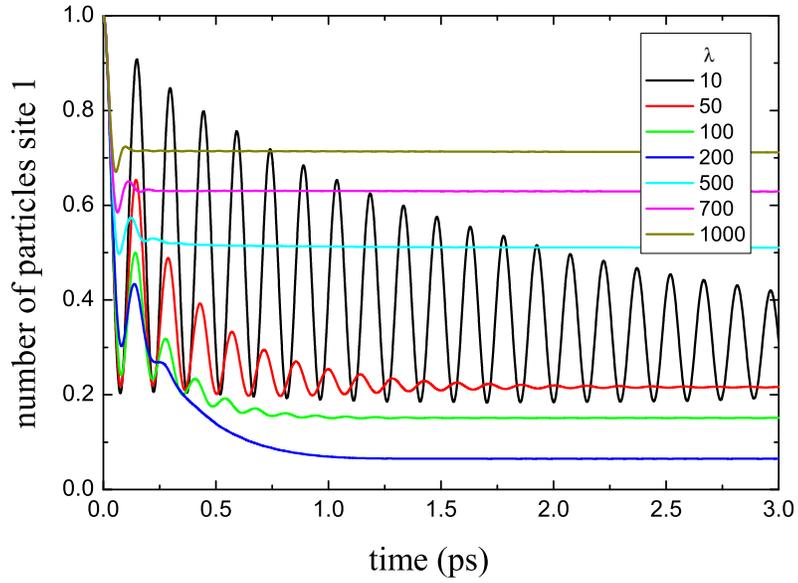}
\caption{Evolutions of the population on site $1$ for the spectral density of Eq. (\ref{jobo}) at $T=0$ K, and various reorganisation energies $\lambda$. Simulation parameters are $J=100\textrm{cm}^{-1},\epsilon_{1}-\epsilon_{2}= 100\textrm{cm}^{-1}$ and $\gamma=53 \textrm{cm}^{-1}$}
\label{fig3}
\end{figure}

Figure (\ref{fig3}) shows the population on site 1 as a function of time for various values of $\lambda$. For $\lambda\leq 100
\mathrm{cm}^{-1}$ we find damped oscillations which persist for at least $1$ ps.
For larger $\lambda$, coherent dynamics are always seen for a few hundred femtoseconds
before the dynamics becomes incoherent, although as $\lambda$ increases the duration of coherent motion becomes shorter. For $\lambda\geq 200
\mathrm{cm}^{-1}$ the incoherent relaxation rate decreases dramatically, and
an increasingly large population is trapped on site $1$
over the timescale of the simulations. This quantum-Zeno-like phenomenon may be related to the well-studied localisation transition
found in Ohmic and sub-Ohmic spin-boson models at $T=0$ K \cite{weiss, leggett, vojta2005quantum,winter09,alvermann09,chin06}. This is a non-perturbative feature of the dynamics, and similar dynamics have also recently been
observed in NRG and NPI studies of the sub-Ohmic spin-boson model \cite{anders07,nalbach2010ultraslow}.

\subsection{ Other spectral densities}\label{ARspec}
We now demonstrate the versatility of our method w.r.t. the microscopic system-bath interactions by considering a much more complex and structured environmental spectral function taken from a recent study of photosynthetic EET. In Ref \cite{adolphs2006proteins}, Adolphs and Renger use a combination of super-Ohmic densities and a coupling to a single effective high-energy mode to model the environment. In our notation this spectral function can be written as,
\begin{eqnarray}
J(\omega)&=&\frac{2\pi\lambda\,[1000\omega^{5}e^{-\left(\frac{\omega}{\omega_{1}}\right)^{\frac{1}{2}}}+4.3\omega^{5}e^{-\left(\frac{\omega}{\omega_{2}}\right)^{\frac{1}{2}}}]}{9!(1000\omega_{1}^{5}+4.3\omega_{2}^{5})}\nonumber\\
&+&4\pi S_{H}\omega_{H}^{2}\delta(\omega-\omega_{H})\label{JAR},
\end{eqnarray}
where we have kept the relative contributions of the two continuous parts of the spectral density as they are in \cite{adolphs2006proteins}, but have also introduced an overall reorganisation energy $\lambda$ to be used as a free parameter. The coupling to the high-energy mode is fixed, and the parameters of the simulation are $J=100\textrm{cm}^{-1},\epsilon_{1}-\epsilon_{2}= 100\textrm{cm}^{-1},\omega_{1}=0.5 \textrm{cm}^{-1}, \omega_{2}=1.95 \textrm{cm}^{-1}, \omega_{H}=180 \textrm{cm}^{-1}, \omega_{c}=1000 \textrm{cm}^{-1}$ and $S_{H}= 0.22$ \cite{adolphs2006proteins}. With these values the continuous part of $J(\omega)$ extends over a frequency range of about $900 \mathrm{cm}^{-1}$, and $\omega_{H}$ is almost resonant with the energy difference ($ ~224\mathrm{cm}^{-1} $) of the dimer eigenstates of $H_{s}$ as the coupling strength of this mode to a site is $~84 \mathrm{cm}^{-1}$.

The chain transformation and DMRG method offers numerical advantages over some other techniques for spectral functions which contain delta functions or damped resonances, as strong coupling to such modes of the environment do not have to be considered as part of the system Hamiltonian. As we discussed in Section \ref{dis}, discontinuous features in the spectral density simply modify the MOPs of the chain mapping, allowing simulation of an arbitrary number of such discrete mode interactions in the presence of a continuous background without any increase in the complexity of the simulation.  Coupling to undamped modes with frequencies comparable to or smaller than the dimer energies have to be considered as part of the system in approaches like NPI, or if included in the spectral function, they must be artifically damped so that their long-time correlation function decays fast enough to be treated accurately within the finite memory time imposed on these methods.

\begin{figure}
\includegraphics[width=12cm]{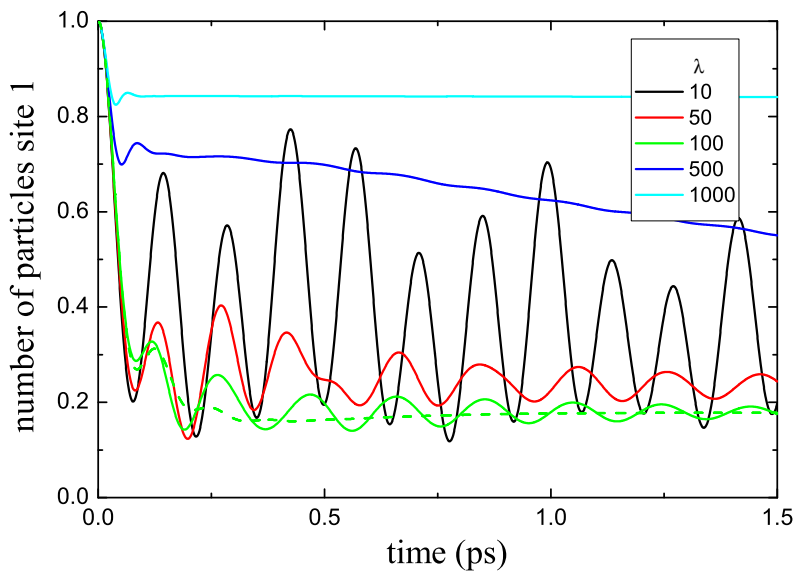}
\caption{Evolutions of the population on site $1$ for the spectral function of Eq. (\ref{JAR}) at various reorganization energies $\lambda$ and $T=0$ K. Dimer parameter are $J=100\textrm{cm}^{-1},\epsilon_{1}-\epsilon_{2}= 100\textrm{cm}^{-1}$.  Dashed line shows how the dynamics when the high-energy mode is decoupled. }
\label{fig4}
\end{figure}

The interaction with the near-resonant oscillator has a pronounced effect on the population dynamics, and Fig.\ref{fig4} shows how this coupling leads to a coherent beating effect which periodically suppresses population oscillations for $\lambda\leq 300 \mathrm{cm}^{-1}$. These coherent multi-frequency effects are a strong sign that even though we treated the discrete mode as part of the environment, the mapping and t-DMRG method accurately treat the quantum coherent interations with this mode. 
In situations where site $2$ might transfer population to another system, such a coherent suppression of oscillations could lead to an enhancement of EET from the dimer to that system. As $\lambda$ increases, the continuous part of the spectral density dominates the dynamics and we observe qualitatively similar behaviour to the dynamics obtained in Fig. \ref{fig3}. We note that the trapping-like dynamics for large $\lambda$ is less severe for this super-Ohmic $J(\omega)$, although the dynamics are still highly non-Markovian for strong coupling.

A particularly striking feature of Fig. \ref{fig4} is that in the regime of optimal EET ($\lambda \sim 100 \mathrm{cm}^{-1}$), the high-energy mode leads to low amplitude oscillations which persist for at least $1.5$ ps. When the high-energy mode is decoupled, coherent oscillations vanish for $\lambda=100\mathrm{cm}^{-1}$ after just $~0.3$ ps. Experimental observation of such persistent undamped oscillations after a fast population transfer could thus indicate the presence of discrete high-energy modes in the environment of PPCs, and could be a useful signature for determining realistic $J(\omega)$s in these complexes~\cite{prior10,chin2010noise}. We also note that broadening the discrete mode by replacing the delta function in Eq. (\ref{JAR}) with an appropriate lineshape for a damped oscillator leads to damping of these long-lasting oscillations (not shown), indicating that these features are induced by the \emph{quantum} nature of the interaction to the resonant discrete mode. Recent experiments on the FMO complex have observed extremely long electronic coherence times of $1-2$ ps, which could be consistent with the effects described above, as vibrational coherences are typically much longer-lasting than electronic coherences.

\section{Conclusions and future developments and applications}\label{conclusions}

\begin{figure}
\includegraphics[width=12cm]{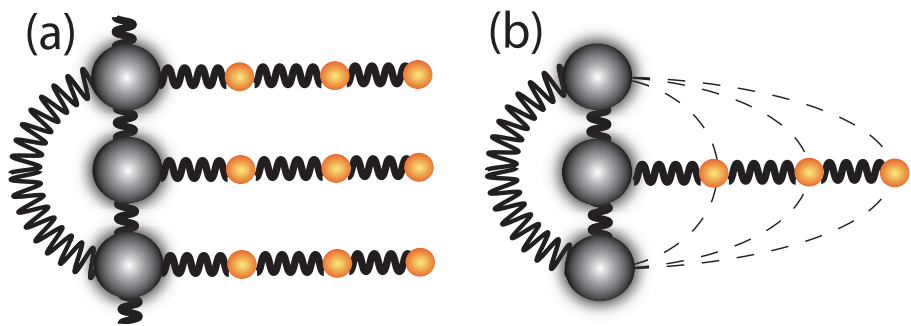}
\caption{(a) A multi-site configuration with independent baths which could in principle be simulated using recent developments in t-DMRG techniques. (b) Multiple sites coupled in a correlated way to a common environment require the treatment of long-range interactions between the sub-system and chain.}
\label{multi}
\end{figure}
In this chapter we have presented the formal development of a mapping technique which converts the standard representation of open-system Hamiltonians into a $1D$ chain Hamiltonian with nearest-neighbour interactions. Using orthogonal poynomials we have found a way to carry out this transformation exactly, and in doing so have rigourously demonstrated a number of hitherto unrecognised universal properties of typical open-system structures. Although this chain mapping is a fascinating subject in its own right, and one that is currently being actively investigated, it also provides a representation which allows the powerful t-DMRG algorithm to be used in simulating open-system dynamics under complex, non-perturbative and structured environmental interactions. Such environments are thought to play an important role in photosynthetic excitation dynamics and the accuracy and versatility of the t-DMRG approach has been illustrated in our numerical examples, where it was discovered that discrete resonances in the spectral function can induce long-lasting coherent dynamics of similar duration to those observed in some PPC complexes. Because this approach simulates the entire many-body wavefunction, it should also allow us to study the dynamical generation of correlations and entanglement between the system and bath, permiting us to explore the ideas of universality and bath reduction schemes presented in Section \ref{universal}. An important practical application of this bath analysis would also be to examine in microscopic detail how vibrational wavepacket dynamics generated by sudden photoexcitation can effect EET dynamics. The microscopic nature of the quantum states leading to the long-lasting electro-vibronic coherences can also be inferred from such an analysis.  

However, detailed simulations of the PPC systems which could be compared to experimental data require a number of technical developments of the method used to produce the results of Prior et al. \cite{prior10}. The most obvious is the need to account for finite temperatures, a problem which have already been resolved with the recent development of mixed-state t-DMRG algorithms \cite{zwolak2004mixed}. Another development is the extension of the method to multi-site networks with independent enviroments. Performing the chain transformation on such a system leads to the Hamitonian structure shown in Fig. (\ref{multi})a. This system can still be treated as as an effectively $1D$ chain with larger local dimensions, allowing standard t-DMRG to be applied. Finally, many current theories about the long-lasting coherence in PPCs invoke the idea that spatial correlations of environmental fluctuations may support long-lasting quantum coherences in these structures. Assuming that the sites couple in different ways to a common environment, the effects of spatial correlations can be investigated. The chain representation of such an open-system is shown in Fig. (\ref{multi})b. Although the number of environmental degrees of freedom to simulate is reduced, one is now faced with having to deal with longer-range interactions between the sites and chain. Extensions of t-DMRG to handle long-range interactions have also recently been developed. Taken together, these developments indicate that extremely efficient, accurate, and completely general algorithms for simulating open-system dynamics have come a step closer.   

We would like to thank J. Prior, A. Rivas, R. Bulla, F. Caruso, F. Caycedo,  J. Almeida, A. Nazir and  A. Datta for discussions on these topics. This work was supported by the Alexander von Humboldt Foundation, the EU STREP projects CORNER, HIP and PICC and the EU Integrated Project QESSENCE.

\end{document}